\

# Generalized TASEP on open chains with a modified injection condition

A.M. Povolotsky[(1)], N.C. Pesheva[(2)], and N.Zh. Bunzarova[(2)],

[(1)]*Bogoliubov Laboratory of Theoretical Physics,
Joint Institute for Nuclear Research, Dubna 141980, Moscow reg., Russia*
[(2)]*Institute of Mechanics, Bulgarian Academy of Sciences,
Acad. G. Bonchev St., block 4, Sofia 1113, Bulgaria*

**Abstract**

We report here our preliminary results in the study of a new version of the generalized Totally Asymmetric Simple Exclusion process (gTASEP) on open tracks. In the gTASEP an additional interaction between the particles is considered, besides the hard-core exclusion, which exists in the standard TASEP. It is modelled by the introduction of a second hopping probability $p_m$ for the particles, belonging to the same cluster, in addition to the standard hopping probability $p$, which now applies only for single particles and the head (rightmost) particle of a cluster. We briefly describe how one can obtain analytically solvable version of gTASEP by modifying the left (injection) boundary condition. Short comparison is made with the previously studied version of gTASEP on open tracks.

## I.  Introduction

The one-dimensional (Totally) Asymmetric Simple Exclusion process ((T)ASEP) is an important model for understanding many non-equilibrium systems. It was first introduced to model kinetics of protein synthesis [1], but later was used to model other non-equilibrium systems as well, e.g., vehicular traffic flow [2-4], biological transport [5] (see, e.g., Fig. 1), forced motion of colloids in narrow channels [6-8] etc.

Its importance stems from the fact that it is one of the few examples of exactly solvable non-equilibrium models. It is also one of the simplest models of self-driven many-particle systems with particle conserving stochastic dynamics, which exhibits non-trivial behavior, i.e., it undergoes phase transitions in one dimension (in the plane of particle input-output rates ($\alpha$, $\beta$)). TASEP is the extremely asymmetric version of ASEP, when particles are allowed to move in one direction only. The model is defined on an open network, in one dimension, in terms of discrete-time, discrete-space stochastic dynamics of hard-core particles.

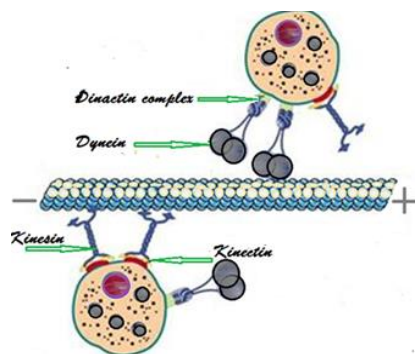

**Fig. 1.**  Illustration of intracellular transport of molecular motors moving along a protofilament. Kinesins move towards the plus end of the microtubule, while dyneins move in the opposite direction.

A comprehensive presentation of traffic flow dynamics and modeling can be found in [9,10].

In the last years the research on non-equilibrium models is intensifying, however, the goal is now to study model systems, which take into consideration more realistic features of real systems. The focus here is on one such model – the generalized TASEP (gTASEP) [11-13].

We report here our preliminary results in the study of a new version of the generalized TASEP (described in more detail in the next section) on open tracks. In the gTASEP an additional interaction between the particles is considered besides the hard-core exclusion in the standard TASEP. It is modeled by the introduction of a second hopping probability for the particles, belonging to the same cluster, in addition to the standard hopping probability, which now applies only for single particles and the head (rightmost) particle of a cluster. We briefly describe how one can obtain analytically solvable version of gTASEP by modifying the left (injection) boundary condition in section III. Short comparison is made in Section IV with the previously studied version of gTASEP on open tracks [13-16].

## II. The model

We focus our study here on the stationary states of the gTASEP with open boundaries, which is an one-dimensional system of hard-core particles of length $L$, obeying a stochastic discrete-time discrete-space kinetics (see Fig. 2). The model is basically the ordinary TASEP with backward ordered sequential update (BSU) [17] with special kinematic interaction, modeled by a second modified hopping probability, $p_m$, for particles in a cluster in addition to the standard hopping probability $p$.

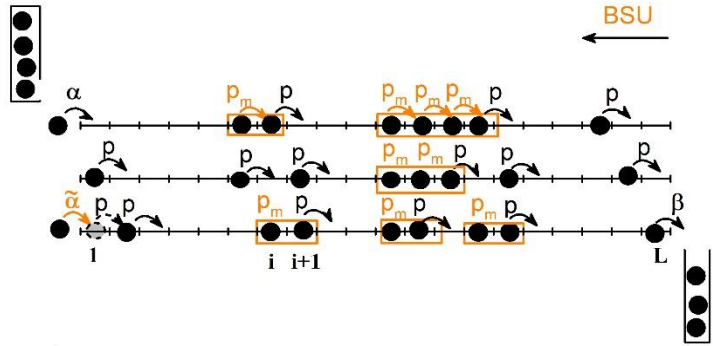

**Fig. 2.** Schematic image of the generalized TASEP with open boundaries (with input-output rates ($\alpha$, $\beta$)), where besides the standard hopping probability $p$, a second hopping probability $p_m$ for the particles in a cluster is introduced. The update rules for the system time evolution are illustrated by showing the change in the system configuration in three consecutive time steps in the case when $p < p_m < 1$ (attraction interaction between the particles in a cluster), when fragmentation of clusters is allowed in the system.

The model is considered a generalization of ordinary TASEP, since for the limiting cases of $p_m = 0$ and $p_m = p$, it reduces to the known and well-studied cases of TASEP with parallel update (PU) [18,19] and TASEP with BSU [17].

The system evolves according to the following rules: If site $i = 1$ was empty at the beginning of the current update, a particle enters the system with probability $\alpha$, or site $i = 1$ remains empty with probability $1 - \alpha$; If site $i = 1$ was occupied at the beginning of the current moment of time, but became empty under its current update, then a particle enters the system with a modified probability

(1) $$\tilde{\alpha} = \min\{\alpha\, p_m/p, 1\}.$$

At the last site $i = L$ of the chain particles are ejected with probability $\beta$. The update sweeps consecutively through the whole chain, starting from the last site $i = L$ (i.e., BSU is used). The left (injection) boundary condition is modified in appropriate manner (it was suggested in [20,21] and independently in [13] to ensure consistency with the update rules in the bulk and also to ensure continuous transition to TASEP with PU (when $p_m = 0$) and TASEP with BSU (when $p_m = p$).

The additional parameter $p_m$ enhances clustering of particles when $p < p_m \leq 1$. This variant of gTASEP (we call it shortly version 1 (V1)) was studied actively in [13-16,22].

However, it appears that one can define the modified injection probability in a different manner, which allows to obtain analytical solution for the steady state probability distribution for gTASEP on open tracks:

$$(2) \qquad \tilde{\alpha} = \frac{\alpha \, p_m (1-p)}{\alpha(p_m - p) + p(1-p_m)}$$

This variant of gTASEP on open chains we call gTASEP V2. This version of gTASEP reduces too to the TASEP with PU when $p_m = 0$ and to the TASEP with BSU when $p_m = p$. How one can arrive at this modified injection probability is shortly accounted for in the next section.

We are interested in the large time behavior of this system on open tracks in the whole range of the parameter $p_m$ controlling the interaction.

### III. Analytically solvable version of gTASEP on open tracks

In the start of this Section let us shortly recall that the matrix product representation of the steady-state probability distribution is a major accomplishment in the methods for solving TASEP on open chains [23]. Different variants of this approach, which became known as the Matrix Product Ansatz (MPA), were suggested and used to obtain exact solutions for the stationary states of TASEP and ASEP under several types of discrete-time stochastic dynamics: sublattice-parallel update [24,25], forward-ordered and backward-ordered sequential updates [17,26], and fully parallel update (simultaneous updating of all sites) [18,19].

Our goal here is to study the gTASEP in the finite case, governed by adding and removing particles at the boundaries. In general, the stationary states of such systems have a complicated structure and in some solvable cases can be found in the form of the matrix product ansatz. This, however, is not an easy task and requires a significant effort to construct the representations of the algebra, corresponding to a given dynamics. It is known, however, that for a special relation between the in- and out-flows the representation of the matrix product algebra trivializes. This happens, when the stationary state on a segment looks like that of a segment in the translation invariant infinite system. Though up to date, the matrix product solution of gTASEP is not known (except for the specific cases of TASEP with PU and TASEP with BSU), one still can ask is it possible to tune the boundary flows so that the finite system looks like a part of the infinite one. To prepare such an infinite system we start from the system on a ring (a finite periodic system with $L$ sites) [12, 27,28]. Starting with the MPA form of the stationary state for the system on a ring one can find the current and the density in the infinite system. From there, involving some further work, one can evaluate the values of probabilities for a particle to enter and exit the system in gTASEP on a segment with open ends, which would assure that the system on a segment looks exactly like a part of the infinite translation invariant system, thus one arrives at the injection boundary condition – Eq. (2). We also obtain the relation

$$(3) \qquad (1 - \beta)(1 - \alpha) = 1 - p,$$

which was already obtained before for the cases of TASEP with PU and BSU, where it defines the regime in which the representation of the algebra used to construct the matrix product stationary state trivializes. Remarkably, this relation holds true also for arbitrary values of $p_m$.

More details of the analytical derivation will be presented in a forthcoming paper [29], which is currently under preparation.

The Monte Carlo simulation results, made for gTASEP V2, are in agreement with the theoretical conclusions, as can be seen from the results, presented in the next section.

## IV. Results and concise account of the differences between gTASEP-V1 and gTASEP-V2

As could be expected the modified version of gTASEP (gTASEP V2) has a phase diagram, consisting of three phases – low density (LD), high density (HD) and maximal current (MC) as in the standard TASEP, and similar to these of the gTASEP V1. However, the two versions have many differences as well. We shortly note some of the observed differences here, some of which are illustrated in Figs. 3,4 below.

In order to confirm the theoretically derived expressions we performed also numerical simulations of the gTASEP V2. Our Monte Carlo simulations of the gTASEP with V2 open boundary conditions confirm the theoretically obtained phase diagram. In what follows we present results, obtained for a system of L=200 sites, $N_{ex}$=21 (time sequence NT=$2^{21}$), $N_{ens}$=100 (ensemble averaging), $N_{tau}$=800 000 (number of time steps omitted so that the system can reach stationary state).

The results (shown in the figures below) for gTASEP V1 are obtained by numerical simulations, while the results for gTASEP V2 are obtained analytically and some results (shown by solid squares) are obtained also numerically in order to verify the analytical calculations.

We show first the density distributions in gTASEP V1 and V2 at $p$ =0.6 and $p_m$ =0.7 at chosen points in the phase space in Fig. 3 (a) and the respective phase diagrams in Fig. 3 (b). The density distributions have the typical shapes, characteristic of the LD, HD and the MC phases and differ slightly for both models.

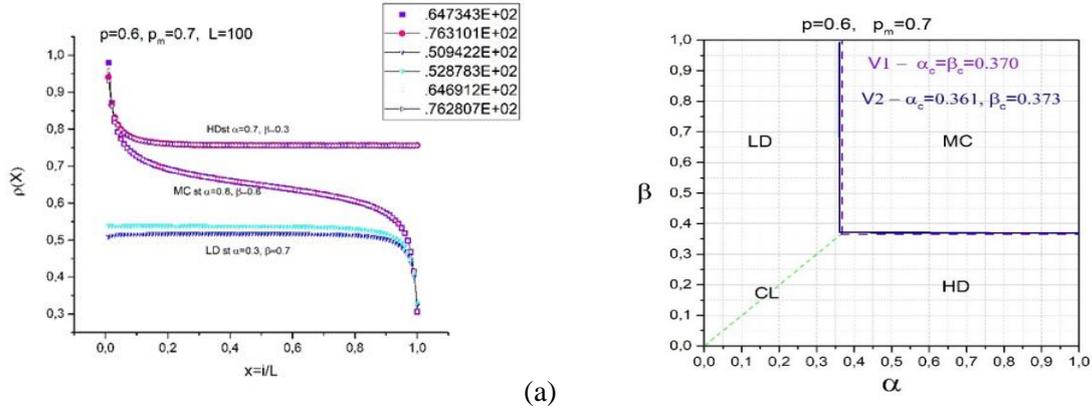

**Fig. 3.** (a) Density distributions along the system in the two versions of gTASEP – V1 and V2, when $p$ =0.6 and $p_m$ =0.7 at chosen points in the phase space, which belong to the LD, HD and MC phases for both models. The results for gTASEP V1 are shown by solid symbols and for gTASEP V2 by empty symbols; (b) The phase diagrams (superimposed) for gTASEP V1 and gTASEP V2 at $p$ =0.6 and $p_m$ =0.7.
One has for gTASEP V1 $\alpha_c = \beta_c = 0.373$ and $\alpha_c \approx 0.3617$, $\beta_c = 0.3733$ for gTASEP V2.

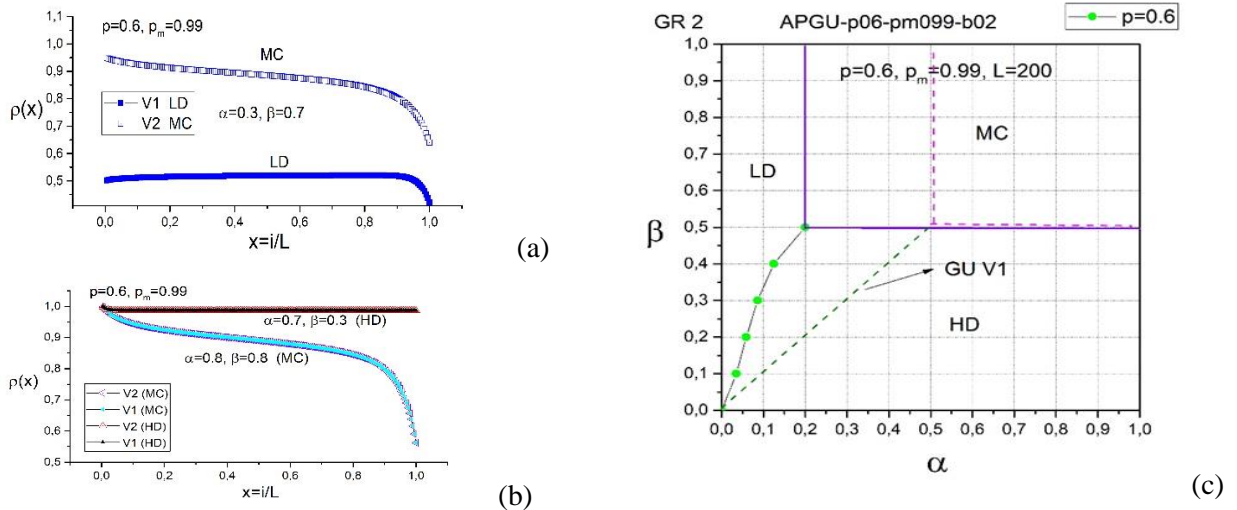

**Fig. 4.** Density distributions along the system in the two versions of gTASEP – V1 and V2, when $p$ =0.6 and $p_m$=0.99: **(a)** at a point $\alpha$=0.3, $\beta$=0.7 (which is in the LD for V1 and in MC phase for V2); **(b)** at point $\alpha$=0.7, $\beta$=0.3 (which is in HD for both V1 and V2) and at $\alpha$=0.8, $\beta$=0.8 (which is in MC p[hase for both V1 and V2). The results for gTASEP V1 are shown by solid symbols and for gTASEP V2 by empty symbols.
**(c)** Phase diagrams for GTASEP V2 and V1 OBC (superimposed), when $p$=0.6, $p_m$=0.99. For gTASEP V1 one has: $\alpha_c$ (0.6; 0.99) = $\beta_c$ (0.6; 0.99) = 0.502, while for the gTASEP V2 one has $\alpha_c$ (0.6; 0.99)=0.211 (0.210) and $\beta_c$ (0.6; 0.99) = 0.493 (0.500). The coexistence line (CL) between the LD and HD phases in the phase diagram of gTASEP V2 is not a straight line;

One of the first observations is that the difference between the phase diagrams of the two versions of gTASEP with OBCs seems to increase with the increase of $p_m$ at fixed $p$. This can be seen by looking at the phase diagrams, presented in Fig. 3 at $p$ =0.6, $p_m$. =0.7 and in Fig. 4, when $p$ =0.6, $p_m$=0.99.

In Fig. 4 we show the phase diagram for gTASEP V2 at $p$=0.6, $p_m$=0.99 with a coexistence line (CL). One of the essential differences is that the CL is not a straight line. The phase diagram of the gTASEP V1 has the same basic structure as that of the standard TASEP, i.e., the CL is a straight line, however, the triple point depends also on the values of the additional parameter $p_m$ [13-16]. We get that the tricritical point for gTASEP V2 is $\alpha_c(0.6,0.99) \approx 0.211$, $\beta_c(0.6,0.99) \approx 0.493$, while for gTASEP V1 we have $\alpha_c$ (0.6; 0.99) = $\beta_c$ (0.6; 0.99) = 0.502 [16,22].

The general topology of the phase diagram of gTASEP V2 remains similar to the one of the ordinary TASEP, it appears though, that when changing the boundary condition at the left end (the particle injection point) of gTASEP leads to a change of the phase diagram expressed in an increase of the area of MC phase and change of the CL from a straight line to a curved (upward) line. This increase is realized by shift of $\alpha_c$ to lower values of α, while $\beta_c$ remains almost a constant. It appears that when $p_m < p$ (repulsion interaction) the CL is (very slightly) curved downward.

We find that both versions of gTASEP with OBCs have $(p, p_m)$-dependent triple point $(\alpha_c, \beta_c)$ and that the main effect of increasing the modified hopping probability $p_m$ is the increase of the MC phase in gTASEP V2 with respect to gTASEP V1. For the shown values of the parameters, the bulk density in the gTASEP V2 is bigger than that in the gTASEP-V1. More work is needed for obtaining a more complete picture of the differences between the two versions.

### V. Conclusion

Most of the real world phenomena and systems are non-equilibrium (in this number are the living systems). They are exchanging matter and/or energy with their environment, having non-trivial currents. Unfortunately, the current understanding of non-equilibrium systems is still behind our understanding of equilibrium systems. In the last 35 40 years research on non-equilibrium models is very active, but more recently the objective has shifted towards more realistic non-equilibrium systems. After obtaining the exact solutions for standard TASEP with different types of updates, now the goal is going further and study and solve models (including different variants of TASEP and TASEP-like models), having additional interaction between the particles, e.g., like the gTASEP or different non-homogeneities [30-32] etc.

The focus here is on one such model – the generalized TASEP (gTASEP). We have suggested how to modify the left (injection) boundary condition in order for the stationary state on a segment to look like that on a segment in the translation invariant infinite system, In this way one finds analytical solution for $(\alpha_c, \beta_c)$ for the system with OBCs.

In the equilibrium statistical mechanics, studies of simple models was very instructive. One may hope that similarly in the non-equilibrium case the study of simple models (like TASEP and its different variants) can be very helpful (instructive) as well.

**Acknowledgements**

Partial financial support by the Bulgarian MES through Grant No. D01-221/03.12.2018 for NCDSC—part of the Bulgarian National Roadmap on RIs, and Grant No. DO1-223/22.10.2021 are thankfully acknowledged.